\documentclass{article}
\usepackage{spconf,amsmath,graphicx}

\usepackage{diagbox}
\usepackage{multirow}
\usepackage{tabularx}
\usepackage{array}
\usepackage{adjustbox}
\usepackage{subcaption}
 \usepackage{hyperref}

\title{From Division to Decision: Leveraging Temporal Cell-Stage Segmentation for Embryo Transferability Prediction}

\name{}
\address{Author Affiliation(s)}

\begin{document}
\twoauthors
  {Y. Hachani$^1$, P. Bouthemy$^1$, E. Fromont$^{1,2}$ 
 }
	{$^1$Inria center at Rennes University, \\
     $^{2}$University of Rennes, IRISA\\
	France}
 {V. Duranthon$^{3}$, L. Laffont$^{3}$, A. P. Reis$^{4,3}$  
  }
{$^{3}$Paris-Saclay University, UVSQ, INRAE, BREED,\\
 $^{4}$The National Veterinary School of Alfort \\
    France}

\maketitle


\begin{abstract}
Accurate selection of bovine embryos is a challenging task, as current practice relies on a single expert assessment on the seventh day after insemination, resulting in high rates of pregnancy loss. Time-lapse videomicroscopy provides detailed information on early development, but is difficult to exploit because of complex motion patterns and time-consuming analysis. We propose TransFACT, a transformer-based framework for modeling early developmental stages and embryo transferability using 2D time-lapse videos from the first four days of development. 
TransFACT combines frame-level temporal features with stage-level representations, using developmental stages as auxiliary supervision to predict transferability on day four. Our experiments demonstrate that TransFACT, by leveraging an existing method designed for action recognition, achieves superior performance than its competitor in predicting embryo transferability. 
\end{abstract}

\section{Introduction}

\textit{In vitro} production (IVP) of bovine embryos is an assisted reproductive technique (ART) that is used to improve genetic gain in farm animals. An IVP embryo is considered transferable if it reaches the blastocyst stage, which is suitable for transfer to a cow uterus. The current method of embryo selection \cite{IETS_2025} involves making a single observation on day seven after insemination (7 DPI) to determine whether the embryo has reached the blastocyst stage. However, this method is inaccurate and results in 70\% of pregnancy loss after transfer. The single observation at 7 DPI provides an incomplete view of the early development of bovine embryos and hide anomalies that occurred earlier. Additionally, it is conceivable to predict transferability as early as 4 DPI. Firstly, an earlier transfer would limit the duration of culture under sub-optimal conditions. Secondly, embryologists must know sooner which embryos are transferable or not for further studies on the causes of miscarriage. 

Analyzing early embryonic development significantly improves the prediction of blastocyst formation, quality, and viability \cite{reis_AETE_2018}. Normal mammalian embryo development is characterized by a series of mitoses, also referred to as \emph{cleavage}, in which each cell divides to form two daughter cells. However, some anomalies may occur, such as direct cleavage, where one mother cell produces more than two daughter cells, reverse cleavage, where two daughter cells merge to form one, cellular death, or, sometimes, development may be arrested. In this paper, the stages of embryo development are defined by the number of visible cells, as well as by the events of cleavage or development arrest.

Monitoring the early development of bovine embryos with advanced microscopy devices requires removing embryos from the incubator. This would disrupt temperature and pH conditions and can negatively affect development. videomicroscopy inside the incubator captures images at short and regular intervals, but it involves a simple miniaturized microscopy technique. Nevertheless, it allows embryologists to identify developmental stages almost in real time. It does so without disturbing the culture conditions required for normal development. It also remains compatible with embryo transfer to establish pregnancy. However, manually determining developmental stages is time-consuming. This limits how laboratories can use the information obtained through videomicroscopy to assess the transferability of embryos. Therefore, automating this process is highly desirable.

Our objective is to predict the transferability of embryos at four days at most (4 DPI). This involves taking 2D time-lapse videos as input and integrating development information into the transferability prediction. Our approach is inspired by the FACT model \cite{lu_CVPR_2024}. It involves training a transformer to jointly predict successive developmental stages and the transferability outcome. We name our model TransFACT. It enables to incorporate information about embryonic development into transferability modeling. We formulate this problem as one of a binary supervised classification problem. An embryo is categorized as either transferable ($T$ class) or not transferable ($NT$ class). 

\begin{figure}[tbh!]
    \centering
    \includegraphics[width=\columnwidth, height=0.25\textheight, keepaspectratio]{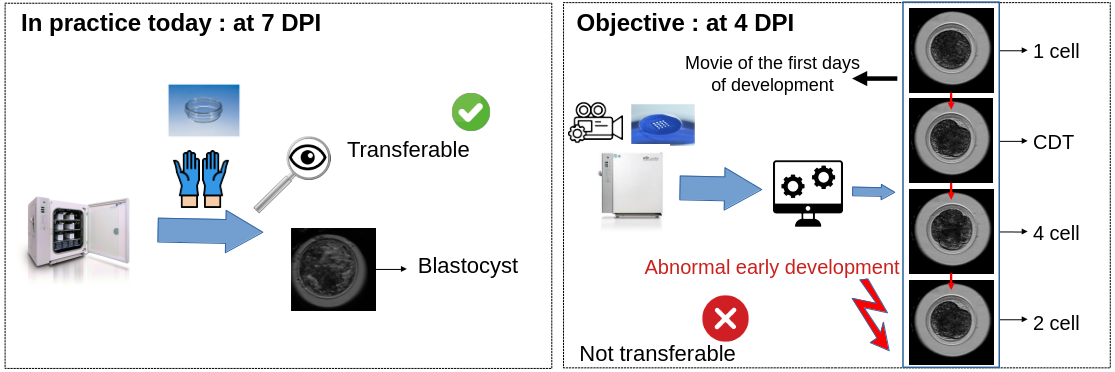}
    \caption{The left box shows the standard workflow in which embryos are manually assessed at 7 DPI. It also provides an example of an embryo of class $T$. The right box illustrates our approach that uses developmental information from time-lapse videos to predict transferability at 4 DPI. The associated example depicts an embryo of class $NT$ with abnormal early development.}
    \label{objectiveTransFACT}
\end{figure}

There are two main reasons why this problem remains challenging, even in a supervised setting: the complex appearance and motion of embryo videos, and the ambiguity between the $T$ and $NT$ classes. First, as shown in Fig.\ref{objectiveTransFACT}, microscopy videos tend to exhibit low contrast and noise, as well as entangled motion. 
Second, high intra-class variability and low inter-class distance make early transferability classification particularly challenging.

The paper is organized as follows. Section \ref{rw} covers related work. Section \ref{method} introduces our TransFACT model. Section \ref{results} presents the experimental results and comparisons. We conclude in Section \ref{conclusion}.

\section{Related work}
\label{rw}

\subsection{Embryo videomicroscopy analysis}
In recent years, the growing importance of ART has led to several studies applying deep learning to human embryo videomicroscopy to predict viability. For example, the authors of \cite{berntsen_Plos_2022} combine an Inflated 3D (I3D) \cite{i3D_CVPR_2017} CNN with a recurrent network, while the ones of \cite{kragh_TMI_2022} use semi-supervised transfer learning based on temporal cycle-consistency.

Embryologists assess embryo quality by analyzing developmental stages and detecting anomalies, which has motivated work on stage prediction. Models such as ESOD \cite{esod_MICCAI_2021} and CNNs-CRF \cite{cnn-crf_MICCAI_2021} incorporate temporal modeling and developmental constraints, while more recent approaches like EmbryosFormer \cite{embryosformer_WACV_2023} rely on transformers. Importantly, the work of \cite{kalyani_SR_2024} shows that explicitly modeling embryonic stages can improve viability prediction.

Little work has been done on bovine embryos. They are more difficult to study than human embryos, because their cells are darker \cite{Hirotada2001}, making tasks such as cell counting more challenging. The authors of \cite{sfr_ICIP_2024} designed a 3D-CNN called SFR that comprises three paths and the focal loss, to predict transferability directly from videos. However, their approach does not incorporate the informative embryonic developmental stages. In addition, they used a 3D-CNN architecture, while the current state-of-the-art performance is achieved with transformer-based models. Notably, SFR is the only existing method specifically designed for bovine embryo transferability prediction.

\subsection{Video action segmentation}
Embryonic development could also be viewed as a sequence of actions. Therefore, we can draw a parallel between our auxiliary task of classifying developmental stages and the task of recognizing and segmenting actions in videos. We now provide a brief overview of deep learning-based methods for action segmentation in videos.

Early approaches to action segmentation are dominated by frame-based models such as MSTCN \cite{farha_CVPR_2019} and the encoder-decoder C2F-TCN \cite{singhania_TPAMI_2023}, which utilizes multi-stage temporal convolutions for frame-level refinement. Although efficient, these methods struggle with capturing long-range temporal dependencies and often suffer from over-segmentation. 
In parallel to convolutional approaches, recurrent architectures, such as BiLSTMs \cite{singh_CVPR_2016} and RNN–HMM hybrids \cite{kuehne_TPAMI_2020} model temporal dependencies sequentially, but prove unscalable and ineffective for long videos \cite{lea_CVPR_2017}.
More recently, two-stage frameworks, such as UVAST \cite{behrmann_ECCV_2022} or UFSA \cite{tran_WACV_2024}, separate frame and action modeling. This improves action-level representation, but typically lacks bidirectional interaction between the respective features, and often required costly post-processing.
The FACT \cite{lu_CVPR_2024} model addresses these limitations through a dual-branch architecture that performs temporal modeling at both the frame and action levels. The frame branch is based on dilated temporal convolutions that preserve fine-grained temporal details. The action branch is built on transformer-based learnable action tokens that capture long-range dependencies and high-level action semantics. The two branches are connected through bidirectional cross-attention within successive update blocks to refine frame and action tokens features. This framework achieved state-of-the-art performance, while offering low computational cost.

\section{Model description}
\label{method}
The dynamics of embryo development has an impact on its transferability. Therefore, it is important to consider the successive stages undergone by the embryo to assess its suitability for transfer.
The stages of embryo development span multiple consecutive frames and are defined by temporally consistent visual patterns rather than isolated images. Each stable stage corresponds to a morphological configuration over a temporal window and typically ends with a dynamic transition stage, such as cell cleavage. We categorize the embryo development stages into 11 classes: nine classical ones corresponding to cell counts (1 to 9+ cells), and two additional ones for cell cleavage events and developmental arrests.
From a video analysis perspective, this temporal structure can be seen as a sequence of actions, each of which is characterized by a beginning, a duration, and a transition to the next state. Consequently, our approach could benefit from action recognition and segmentation strategies.

We present our proposed TransFACT architecture below, along with an optional Motion History Images (MHI) component to account for the critical motion events during the embryo's development.

\subsection{Network architecture}

Our model TransFACT must encompass two objectives of different but complementary nature: \textit{division} by segmenting the successive stages of development, and \textit{decision} by predicting the embryo transferability. To achieve the division goal, we were inspired by the recent FACT framework \cite{lu_CVPR_2024}. Indeed, its temporal modeling strategy aligned well with embryonic development. However, we need to augment this framework with the prediction of transferability, a label concerning the whole video, to achieve the decision goal.

Modeling stage-level temporal context alongside frame-level features ensures temporal consistency and enhances the management of transitions between developmental stages, especially when frame-level visual cues are ambiguous.
Embryonic development stages act as semantic temporal units that organize the visual evolution of the embryo over time.

We adapt the FACT architecture by substituting a stage branch for the action branch, in which learnable stage tokens represent embryonic developmental stages and their temporal extent. These stage representations interact with frame-level features to provide a global developmental context. This integration enables us to disambiguate local predictions and enforces the succession of biologically consistent stages. Furthermore, we introduce a global classification head informed by frame and stage-level features to predict transferability, as summarized in Fig. \ref{fig:FACT-architecture}. 
In line with FACT, we used pre-extracted frame-level features for computational efficiency, allowing TransFACT to be trained end-to-end.

\subsection{Motion history images}

Since the movement visible in the embryo video is rather intricate and difficult to estimate, we turned to motion-related information that is easy to compute and relevant for a static camera: Motion History Images (MHI) \cite{davis_CVPR_1997}.

\begin{figure}[!t]
    \centering
    \includegraphics[width=\linewidth, height=0.5\textheight, keepaspectratio]{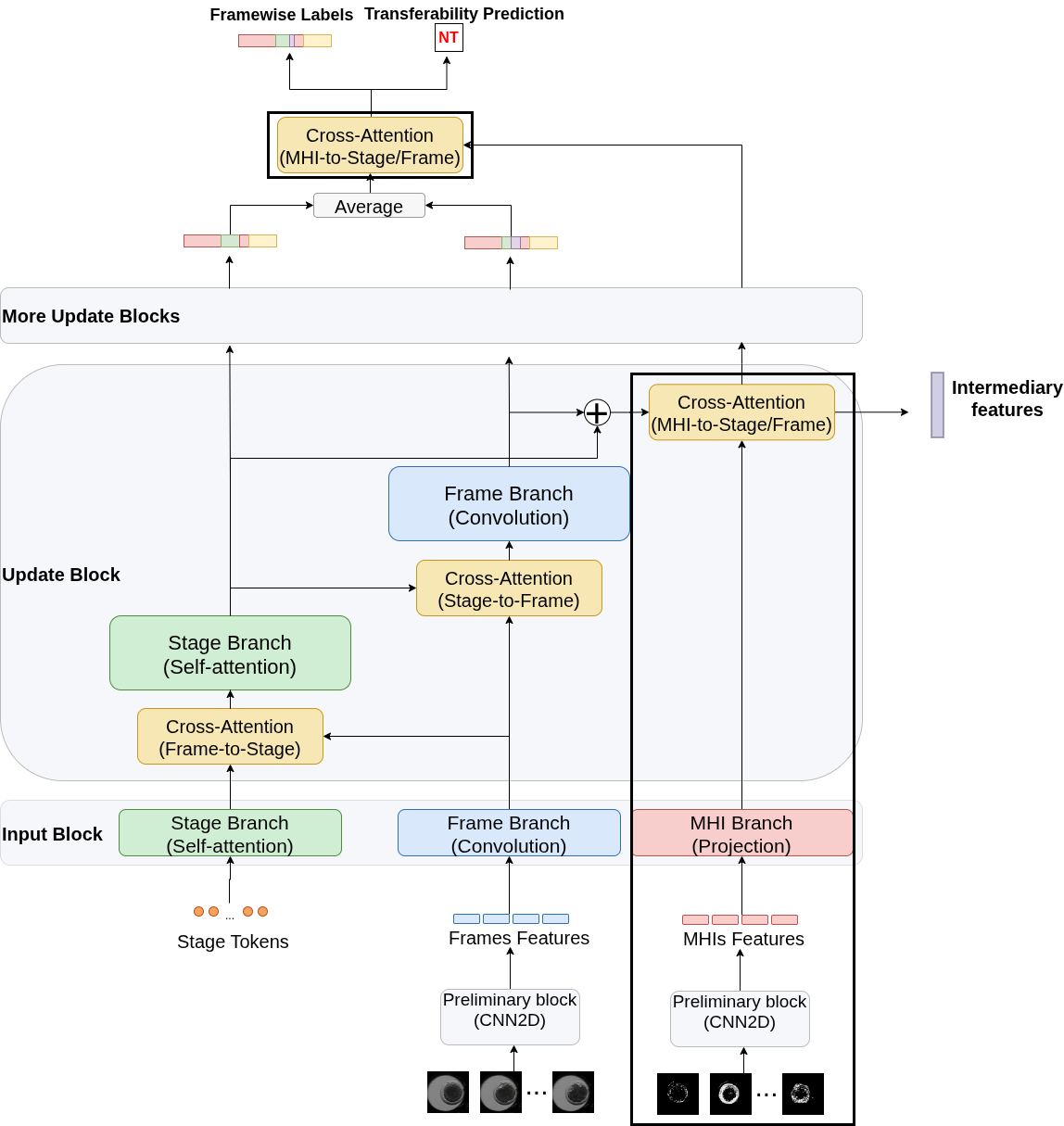}%
    \hfill
    \caption{TransFACT architecture with optional integration of MHIs (additional MHI modules are boxed in bold). The segmentation part is inspired by the FACT model \cite{lu_CVPR_2024}. The framework jointly predicts frame-wise developmental stages and global embryo transferability from time-lapse videos.
    During training, supervision is applied to the final output and the intermediate outputs of each block.
    }
    \label{fig:FACT-architecture}
\end{figure}

MHI maps constitute compact motion information that encodes temporal patterns by aggregating the presence of pixel-wise motion over a sequence of grayscale frames. A MHI map $H_{\tau}(x,y,t)$ at frame $t$ and pixel $(x,y)$ is computed over a temporal window of length $\tau$ using a binary mask given by:
\begin{equation}
D(x,y,t) = 
\begin{cases} 
1 & \text{if } |I(x,y,t) - I(x,y,t-1)| > \theta, \\
0 & \text{otherwise},
\end{cases}\end{equation}
\noindent
where $I(x,y,t)$ is the pixel intensity at $t$, and $\theta$ a threshold. The MHI is then updated as:
\begin{equation}H_{\tau}(x,y,t) = 
\begin{cases} 
\tau  \;\;\;\;\; \text{ if } D(x,y,t) = 1, \\
\max(0,H_{\tau}(x,y,t-1) - 1)  \text{ otherwise}.
\end{cases}\end{equation}
As shown in Fig. \ref{fig:FACT-architecture}, we integrate the MHI features by introducing an additional cross-attention mechanism, which allows adaptive feature fusion. Adding MHI marginally increases FLOPs from 3.03G to 3.31G. 

\begin{table*}[tbh!]
\centering
\begin{adjustbox}{max width=\textwidth}
\begin{tabular}{l|lll|lll|l}
\hline\hline
\diagbox{Modality}{Metric}           & \multicolumn{1}{c}{$P_T$} & \multicolumn{1}{c}{$R_T$} & \multicolumn{1}{c|}{$F1_T$} & \multicolumn{1}{c}{$P_{NT}$} & \multicolumn{1}{c}{$R_{NT}$} & \multicolumn{1}{c|}{$F1_{NT}$} & \multicolumn{1}{c}{Accuracy}  \\ 
\hline
TransFACT with frames                         & $\mathbf{80.11 \pm 2.07}$             & $84.94 \pm 2.59$             & $\mathbf{82.41 \pm 0.82}$     & $85.52 \pm 1.77$              & $\mathbf{80.66 \pm 3.12}$              & $\mathbf{82.97 \pm 1.20}$       & $\mathbf{82.70 \pm 0.9}$   \\
TransFACT with MHIs                           & $76.35 \pm 1.55$             & $81.57 \pm 1.01$             & $78.86 \pm 1.13$               & $82.06 \pm 0.99$     & $76.92 \pm 1.86$              & $79.40 \pm 1.34$               & $79.14 \pm 1.23$             \\
TransFACT with frames+MHIs & $79.20 \pm 1.79$             & $\mathbf{85.66 \pm 4.71}$    & $82.21 \pm 1.29$               & $\mathbf{86.09 \pm 3.63}$     & $79.34 \pm~ 3.43$             & $82.46 \pm 0.57$                & $82.36 \pm 0.69$             \\
\hline
\end{tabular}
\end{adjustbox}
\caption{Performance of our TransFACT model using frame features, MHI, and their combination as input. Prediction, recall and F1-score per class, along with the global accuracy are reported as mean $\pm$ std across five runs with different random seeds.}
\label{tab:mhi-classif}
\end{table*}

\subsection{Loss function}

In the FACT model \cite{lu_CVPR_2024}, the training step jointly supervises the frame-wise prediction and the stage-level representation. We add a prediction head for transferability and a fifth loss term $L_{\text{trans}}$, enabling optimization of this additional objective. The loss function is now defined as: $L = L_{\text{trans}} + L_{\text{frame}} + L_{\text{stage}} + L_{\text{cross-att}} + L_{\text{smooth}}$. 
Let \( v \) be a video of \( T = 300 \) frames and \(N \) segments of consecutive, identically labeled frames. Each segment \( n \) corresponds to a temporal interval \(  T_n \). We set \( S=11 \) developmental stage labels, \( M=60 \) stage tokens, and use \( B=3 \) blocks in the model. Outputs from the frame branch and the stage branch carry a superscript \(f \) and \(s \) respectively. $L_{\text{trans}}$ that ensures the correct transferability classification for the video writes:
\noindent
\begin{equation}L_{\text{trans}} = -\sum_{b=1}^B \log p_{trans}^b(v),\end{equation}
\noindent
where \( p_{trans}^b(v) \) denotes the predicted score of the ground-truth transferability label of video \( v \) at block \( b \). $L_{\text{frame}}$ ensures the correct classification of each frame \cite{lu_CVPR_2024}:
\noindent
\begin{equation}L_{\text{frame}} = -\frac{1}{T}\sum_{b=1}^B\sum_{t=1}^{T} \log p_{dev}^{b,f}(v, t),\end{equation}
\noindent
where \( p_{dev}^{b,f}(v, t) \) denotes the predicted score of the ground-truth label of frame \( t \) of video \( v \) at block \( b \). For the stage branch, the objective is defined by $L_{\text{stage}}$ \cite{lu_CVPR_2024}, which supervises the token outputs:
\begin{equation}
\begin{split}
L_{\text{stage}} = & -\frac{1}{M}\sum_{b=1}^B \bigl[\sum_{n=1}^{N} \log p^{b,s
}_{dev}(\pi^*(v, n), s^v_n) \\
                    & + \sum_{m\in\mathcal{N}_v} \log p^{b,s}_{dev}(m, S+1)\bigr],
\end{split}
\end{equation}
\noindent
where \( \pi^*(v, n) \) denotes the token matched to the segment \( n \) of video \( v \), \( s_n^v \) the ground-truth development label of segment \( n \) of video \( v \), \(\mathcal{N}_v\) the set of tokens not matched to any segment of \( v \), \(S+1\) the null class, and \(p^{b,s}_{dev}(i, j)\) the predicted score that token \(i\) belongs to stage class \(j\) at block \( b \). A cross-attention alignment loss further encourages temporal consistency between frames and their matched stage tokens \cite{lu_CVPR_2024}:
\begin{equation}
    \begin{split}
    L_{\text{cross-att}} = -\frac{1}{T}\sum_{b>1}\sum_{n} \sum_{t \in T_n} [ & \log \Lambda^{b,s}(t,\pi^*(v, n)) \\
                                                                                & + \log\Lambda^{b,f}(t,\pi^*(v, n))],
    \end{split}
\end{equation}
\noindent
with \( \Lambda^{b,s}(t,\pi^*(v, n)) \) and \( \Lambda^{b,f}(t,\pi^*(v, n)) \) denoting the cross-attention weights between a frame \(t\) in segment \(n\) and the matched token \(\pi^*(v, n) \), from stages to frames and from frames to stages, respectively.
Finally, $L_{\text{smooth}}$ mitigates over-segmentation by penalizing abrupt changes (see Appendix of \cite{lu_CVPR_2024}).
Together, these loss terms ensure multi-scale temporal supervision that leverages the joint frame-stage representation.
Following \cite{lu_CVPR_2024}, we set unit weights for $L_{\text{frame}}$, $L_{\text{stage}}$, and $L_{\text{cross-att}}$, and 5 for $L_{\text{smooth}}$. A sensitivity analysis shows that removing the transferability loss degrades performance, while the results stabilize for weights within $[0.5, 4]$. We thus use a weight of 1 for the transferability loss.

\section{Experimental results}
\label{results}

\subsection{Video capture and video dataset}

We use time-lapse videos of bovine embryo development from 1 to 4 DPI, acquired every 15 minutes, producing sequences of 300 grayscale frames of size $256 \times 256$. Each video is annotated by a biologist as transferable ($T$) or non-transferable ($NT$).
The dataset contains 1740 videos, split into 1220/175/345 for training, validation, and testing respectively, with 53\% $NT$ embryos in every subset. We will refer to this dataset as INRAE-Gertrude-DT.

\subsection{Implementation details}

Our TransFACT model has one input block and three update blocks. All configurations were optimized using AdamW 
with a learning rate of $10^{-4}$ and a standard learning rate warm-up schedule. Training was conducted for 50 epochs with a batch size of 32. For evaluation, we selected the model checkpoint from the epoch that achieved the lowest validation loss during supervised training. The input frame and MHI features were generated using two different pre-trained 2D-CNN. We set $\tau=15$ and $\theta=20$ to generate the MHI.

\begin{table*}[tbh!]
\centering
\begin{adjustbox}{max width=\textwidth} 
\begin{tabular}{l|lll|lll|l} 
\hline\hline
\diagbox{Model}{Metric} & \multicolumn{1}{c}{$P_T$} & \multicolumn{1}{c}{$R_T$} & \multicolumn{1}{c|}{$F1_T$} & \multicolumn{1}{c}{$P_{NT}$} & \multicolumn{1}{c}{$R_{NT}$} & \multicolumn{1}{c|}{$F1_{NT}$} & \multicolumn{1}{c}{Accuracy}  \\ 
\hline
TransFACT               & $\mathbf{80.11 \pm 2.07}$     & $\mathbf{84.94 \pm 2.59}$              & $\mathbf{82.41 \pm 0.82}$      & $\mathbf{85.52 \pm 1.77}$                 & $\mathbf{80.66 \pm 3.12}$        & $\mathbf{82.97 \pm 1.20}$          & $\mathbf{82.70 \pm 0.90}$        \\
SFR                     & $66.04 \pm 3.01$              & $73.49 \pm 11.76$             & $69.08 \pm 3.72$                & $74.73 \pm 8.67$                 & $65.33 \pm 9.90$                 & $68.75 \pm 3.79$                   & $69.20 \pm 1.84$                  \\ 
\hline
p-value / Cohen’s d     & 0.063 / 3.03              & 0.125 / 0.989             & 0.063 / 4.18                & 0.188 / 0.767                & 0.063 / 1.33                 & 0.063 / 3.03                   & 0.063 / 6.54                  \\
\hline
\end{tabular}
\end{adjustbox}
\caption{Performance metrics for our TransFACT model and SFR \cite{sfr_ICIP_2024}. Results are presented as mean $\pm$ std across five independent runs with different random seeds. Precision, recall and F1-score are reported for $T$ and $NT$ classes. Statistical significance is assessed using a paired Wilcoxon signed-rank test.}
\label{tab:model-comparison}
\end{table*}

\subsection{MHI integration}
\label{mhi_experiment}
The contribution of motion information was evaluated by comparing frameworks with and without MHI input. Evaluation metrics included overall accuracy and precision ($P$), recall ($R$), and $F_1$-score ($F1$) for the $T$ and $NT$ classes. 
Tab.\ref{tab:mhi-classif} summarizes the results of the different input configurations. The TransFACT model using only the frame features achieves the highest overall $F_1$-scores for both classes ($F1_T = 82.41 \pm 0.82$, $F1_{NT} = 82.97 \pm 1.20$) and the best accuracy ($82.70 \pm 0.90$). The MHI-only model shows the lowest performance across all metrics, which means that the video modality remains the dominant contributor to the performance. However, adding MHI features to frame features with cross-attention may be beneficial. It yields a comparable but slightly lower overall accuracy, but with slightly higher stability. In addition, it achieves the highest recall for the $T$ class ($85.66$) and the highest precision for the $NT$ class ($86.09$). Since transferable embryos are valuable, achieving high recall $R_{T}$ for this class is critical. Moreover, a more stable accuracy is of interest to biologists.

\subsection{Comparative experiments}



We compared our TransFACT method to SFR \cite{sfr_ICIP_2024}. For this comparison, we retained the TransFACT variant based on frame-level features, as it provides the highest overall accuracy. The evaluation followed the same protocol as described in Section \ref{mhi_experiment}. When applied to our dataset, SFR yielded lower performance than reported in \cite{sfr_ICIP_2024}, which is likely due to differences in dataset characteristics.
All results are summarized in Table \ref{tab:model-comparison}. TransFACT achieves the best overall performance by a large margin, with a $F_1$ of $82.41 \pm 0.82$ for the $T$ class and $82.97 \pm 1.20$ for the $NT$ class, and a general accuracy of $82.70 \pm 0.90$. In addition to higher scores, TransFACT exhibits greater stability in all evaluation metrics. In contrast, SFR performs consistently worse, highlighting the effectiveness of the TransFACT architecture for accurate classification.
To support this comparison, we conducted statistical significance tests. Although the results do not reach the 0.05 threshold, they consistently favor TransFACT, with a trend toward significance ($p < 0.1$, except for $R_T$ and $P_{NT}$) and large effect sizes (Cohen’s $d > 0.8$, except for $P_{NT}$), indicating a significant performance gap.

\begin{figure}[tb!]
\centering
\resizebox{0.9\columnwidth}{!}{
        \includegraphics{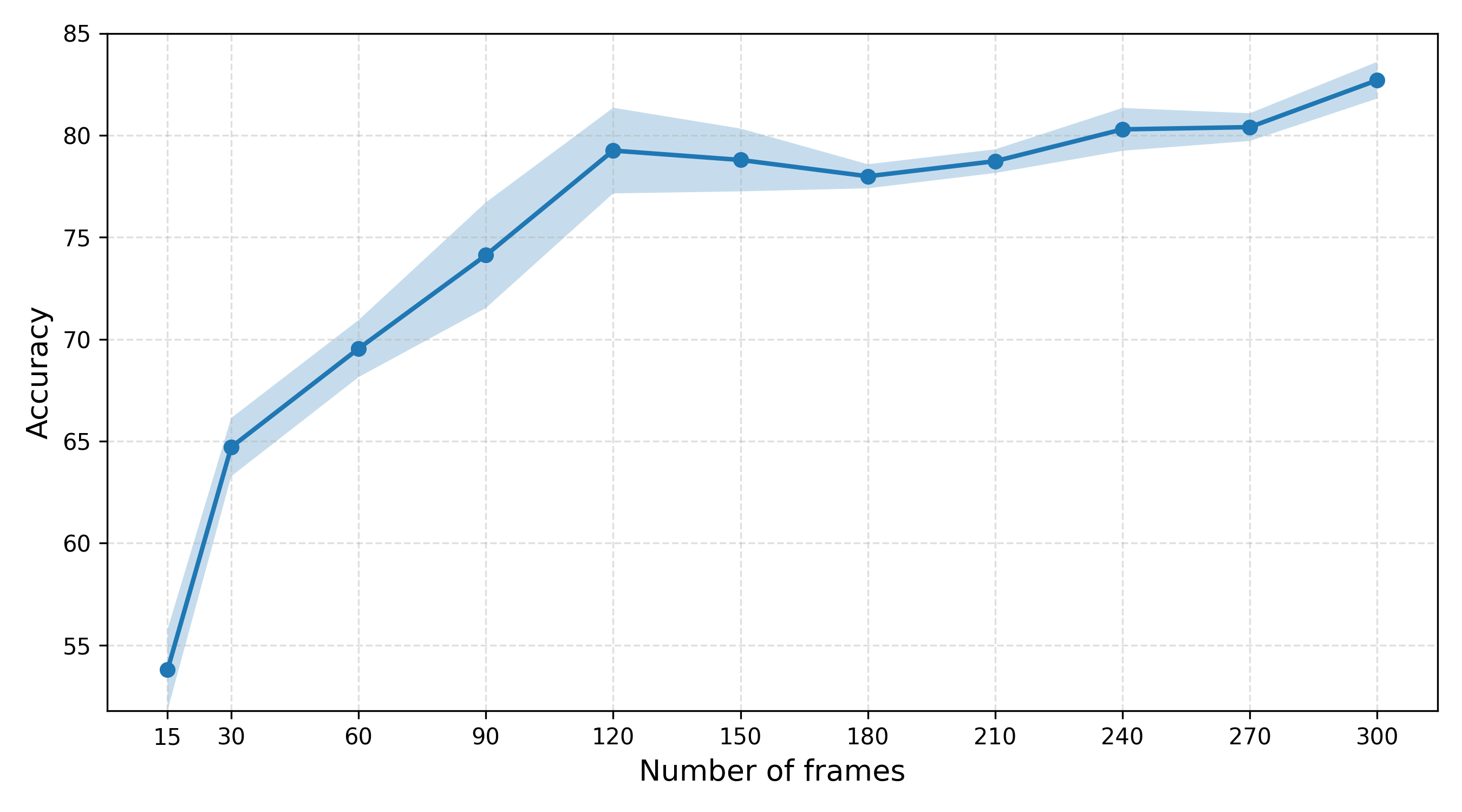}   
}

\caption{Accuracy of TransFACT with respect to input video length (15-300 frames, approximately 1 to 4 days of development). Points show mean accuracy over five runs with different initializations; shaded areas indicate standard deviation.
}
\label{fig:early_prediction}
\end{figure}

We have investigated how the progressive capture of developmental information affects the accuracy of the prediction of embryo transferability. Time-lapse sequences were progressively truncated from the end, and the model was retrained and evaluated at each step to measure performance with reduced temporal context. The results are reported in Fig.\ref{fig:early_prediction}. Accuracy increases as more key developmental events are observed. Starting from only one cell, performance is limited (54\%) and improves with successive cleavages: 65\% after the first cleavage cycle at about frame 30, 70\% after the second cleavage cycle at about frame 60 and 74\% after the third cleavage cycle at about frame 90. A further increase to 79\% occurs when the temporal window includes the start of the LAG phase at about frame 120, a period during which the embryo temporarily pauses, with a final improvement to 82\% upon observing its completion. This last stage allows the detection of embryos that failed to resume development.

We can conclude that the predictive power of TransFACT is acquired cumulatively: the more developmental context the model observes, 
the more accurate the transferability prediction becomes.
Additionally, we observe that transferability can still be predicted with 79\% accuracy using only 120 frames (approximately two days of development), albeit at the cost of increased variability. This would allow biologists to select embryos of interest at a very early stage for subsequent studies on the biological conditions associated with transferability and non-transferability.

\section{Conclusion}
\label{conclusion}

We have presented TransFACT, a transformer-based framework inspired by FACT, for joint modeling of early embryonic development and prediction of bovine embryo transferability using time-lapse videomicroscopy. Our method achieves state-of-the-art performance for the prediction of embryo transferability at 4 DPI by combining frame-wise and stage-level representations and leveraging developmental stages as auxiliary supervision. This method is also stable across experiments. A relatively accurate transferability prediction can even be achieved as early as 2 DPI. This level of performance has been achieved through the appropriate consideration of biologically significant developmental events.

\section{Acknowledgements}
We acknowledge B. Marquant-LeGuienne and CRB-Anim (ANR-11-INBS-0003) for the production of the bovine embryo dataset.
The experiments used Grid'5000 infrastructure
supported by a scientific interest group and hosted by Inria.
\bibliographystyle{IEEEbib}
\bibliography{icip2026}

\end{document}